\documentclass[final]{IEEEtran}
\usepackage{amsmath,amssymb,amsfonts,mathrsfs,bm,amsthm}
\usepackage{amstext}
\usepackage{upgreek}

\usepackage{multicol}
\usepackage{indentfirst}
\usepackage{graphicx}
\usepackage{paralist}
\usepackage{multirow}
\usepackage{tabularx}
\usepackage[noadjust]{cite}

\usepackage{booktabs}
\usepackage{subfigure}
\usepackage{color}
\usepackage{textcomp}

\IEEEoverridecommandlockouts
\allowdisplaybreaks
\usepackage{algorithm,algorithmicx}
\floatname{algorithm}{\bf Algorithm}

\usepackage[noend]{algpseudocode}
\algrenewcommand{\algorithmiccomment}[1]{\hfill // #1}

\makeatletter
\newcommand{\algmargin}{\the\ALG@thistlm}
\makeatother
\algnewcommand{\parState}[1]{\State 
	\parbox[t]{\dimexpr\linewidth-\algmargin}{\strut #1\strut}}

\usepackage{accents}

\allowdisplaybreaks
\definecolor{blue}{rgb}{0,0.24,0.54}
\begin{document}
\title{Deep Learning Aided Routing for Space-Air-Ground Integrated Networks Relying on Real Satellite, Flight,  and Shipping Data} %
\author{{Dong Liu, Jiankang Zhang, Jingjing Cui, Soon-Xin Ng, Robert G. Maunder, and Lajos Hanzo}
	\thanks{D. Liu is with the School of Cyber Science and Technology, Beihang University, Beijing 100191, China. (e-mail: dliu@buaa.edu.cn)}
	\thanks{J. Cui, S. X. Ng, R, G. Maunder, and L. Hanzo are with the School of Electronics and Computer Science, the University of Southampton, Southampton SO17 1BJ, U.K. (email: jingj.cui@soton.ac.uk; sxn@ecs.soton.ac.uk; rm@ecs.soton.ac.uk; lh@ecs.soton.ac.uk).}
	\thanks{J. Zhang is with the Department of Computing and Informatics, Bournemouth University, Bournemouth BH12 5BB, U.K. (e-mail: jzhang3@bournemouth.ac.uk).}
	\thanks{L. Hanzo would like to acknowledge the financial support of the Engineering and Physical Sciences Research Council projects EP/P034284/1 and EP/P003990/1 (COALESCE) as well as of the European Research Council's Advanced Fellow Grant QuantCom (Grant No. 789028).}
}

\maketitle

\begin{abstract}
Current maritime communications mainly rely on satellites having meager transmission resources, hence suffering from poorer performance than modern terrestrial wireless networks. With the growth of transcontinental air traffic, the promising concept of aeronautical ad hoc networking relying on commercial passenger airplanes is potentially capable of enhancing satellite-based maritime communications via air-to-ground and multi-hop air-to-air links. In this article, we conceive space-air-ground integrated networks (SAGINs) for supporting ubiquitous maritime communications, where the low-earth-orbit satellite constellations, passenger airplanes, terrestrial base stations, ships, respectively, serve as the space-, air-, ground- and sea-layer. To meet heterogeneous service requirements, and accommodate the time-varying and self-organizing nature of SAGINs, we propose a deep learning (DL) aided multi-objective routing algorithm, which exploits the quasi-predictable network topology and operates in a distributed manner. Our simulation results based on real satellite, flight, and shipping data in the North Atlantic region show that the integrated network enhances the coverage quality by reducing the end-to-end (E2E) delay and by boosting the E2E throughput as well as improving the path-lifetime. The results demonstrate that our DL-aided multi-objective routing algorithm is capable of achieving near Pareto-optimal performance.
\end{abstract}


\section{Introduction}
Next-generation wireless networks are envisaged to support high-speed, low-latency and high-reliability communications, anywhere and anytime. Yet, more than 70 percent of the Earth surface is covered by oceans and the ever increasing activities scattered across the ocean have created great demand for maritime communications. At the time of writing, shipping mainly relies on satellites for seamless coverage~\cite{wei2021hybrid}. However, due to the wide coverage area of a satellite, the transmission bandwidth allocated to each user device is rather limited, even when high-throughput satellites are relied upon~\cite{vondra2018integration}. Moreover, geosynchronous-equatorial-orbit (GEO) satellites have a high propagation delay of about 120 ms, while their low-Earth-orbit (LEO) counterparts appear above the horizon for short duration and suffer from Doppler-effects.

On the other hand, the number of intercontinental passenger airplanes above the ocean is  significant and there is  an increasing demand for in-flight Internet connectivity. Similar to ships, airplanes also face the same satellite connection limitations as their maritime counterparts. In this context, the compelling concept of aeronautical ad-hoc networking (AANET) was proposed to form a self-configured wireless network via multihop air-to-air (A2A) communication links~\cite{zhang2019aeronautical}. Nevertheless, the coverage of AANETs hinge on the flight density, which fluctuates during a day.

Therefore, it is a nature inspiration to conceive the combination of satellites and airplanes to form a space-air-ground integrated network (SAGIN)~\cite{huang2019airplane,vondra2018integration} for supporting future maritime communications. 
However, due to the inherent characteristics of heterogeneity, self-organization, and time-variability, the design and optimization of SAGINs faces numerous challenges~\cite{liu2018space}. A fundamental one is to design an efficient routing protocol for constructing an appropriate packet routing path at any given time in order to accommodate the high-dynamic network topology.

Recent advances in artificial intelligence have inspired diverse applications in wireless communications~\cite{luong2019applications,liu2020optimizing}, including maritime communications~\cite{yang2020ai}. Taking the routing problem for example, a deep learning (DL) aided routing algorithm was proposed in~\cite{kato2019optimizing} for balancing the traffic in SAGIN. Yet, the network topology was assumed to be static and the global node status had to be known for making routing decisions. To handle the high-dynamic network topology, a deep reinforcement learning (DRL) aided routing algorithm was conceived in~\cite{liu2021deep} for AANETs. By solely relying on local- rather than global-information, the powerful DRL-aided routing advocated is capable of achieving near-optimal end-to-end (E2E) delay. 

Nevertheless, apart from the E2E delay, the overall network performance of SAGIN should be characterized by multiple metrics~\cite{jingjing}, such as the throughput and the path-lifetime to minimize rerouting for maintaining seamless connectivity. The heterogeneous services of SAGINs have different quality of service requirements and it is vital but challenging to strike compelling tradeoffs among the various potentially conflicting optimization objectives. Therefore, instead of finding a particular optimal solution such as the minimum-delay or maximum-throughput route, the ultimate goal is to discover the entire \emph{Pareto front} of all Pareto-optimal routes. Explicitly, for a Pareto-optimal delay-throughput pair, neither of them can be improved without sacrificing the other.

In this article, we propose a satellite-airplane-terrestrial  integrated solution for supporting ubiquitous high-quality maritime communications. In contrast to the satellite-UAV-terrestrial solution proposed in~\cite{li2020enabling}, where the air-layer is only beneficial for local coverage enhancement due to limited energy onboard of unmanned aerial vehicles (UAVs), our air-layer relies on the commercial passenger flights regularly flying across the ocean, hence providing long-duration and wide-range coverage. To solve the challenging multi-component routing-optimization problem high-dynamic networks, we design a DL-aided solution solely relying local- rather than global-information. Simulations are conducted based on real satellite, flight, and shipping data over the North Atlantic ocean. The results show the near-Pareto-optimality of our DL-aided multi-objective routing in terms of reducing  the E2E delay, increasing the E2E throughput, and enhancing the path-lifetime. 
\section{Satellite-Airplane-Terrestrial Integrated Architecture}
In this section, we introduce the  multi-layer network architecture conceived for supporting maritime communications, compare the characteristics of different layers, and formulate the performance metrics. 
\subsection{Multi-Layer Network Architecture}
\begin{figure}[!htb]
	\centering
	\includegraphics[width=0.47\textwidth]{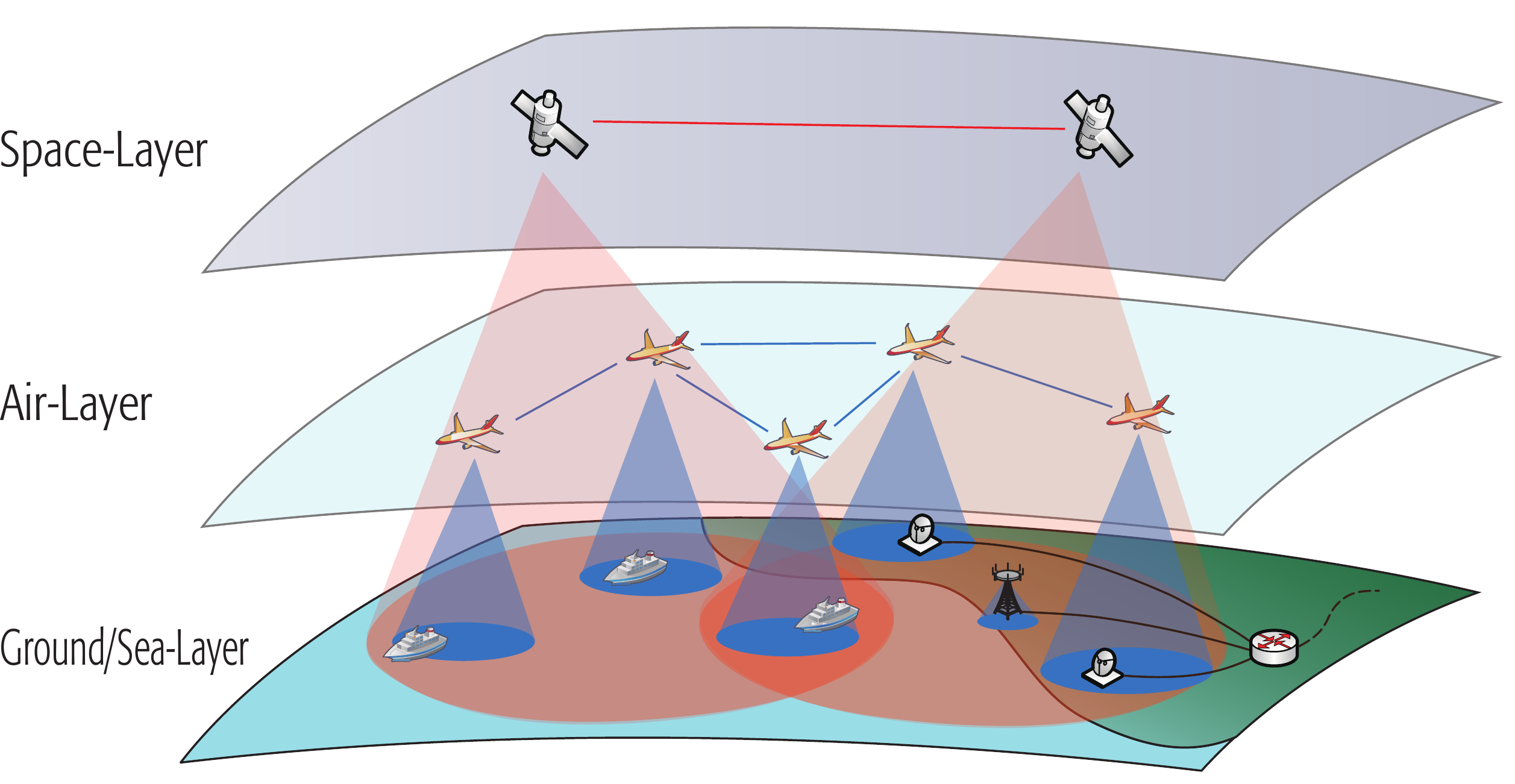}
	\caption{Multi-Layer Network Architecture}
	\label{fig:layer}
\end{figure}

The SAGIN is composed by three layers, namely the ground/sea-layer, the air-layer, and the space-layer, as shown in Fig.~\ref{fig:layer}.

\subsubsection{Ground/Sea-Layer} 
The ground/sea-layer includes ships, on-shore base stations (BSs), and satellite/airplane ground station (GS). Each ship seeks a wireless connection to the BS or the GS on land. The on-shore BSs are deployed along the coast, but their coverage range is quite limited, connecting the ships that are densely clustered in the vicinity of ports, harbors, and waterways. The GSs are deployed for connecting the air- as well as space-layer and serve as the gateway for accessing the terrestrial core network. 

\subsubsection{Air-Layer} 
The air-layer is formed by the AANET consisting of commercial passenger airplanes. Each airplane is able to establish air-to-ground/sea (A2G) links with GSs or ships, as well as establish A2A links with nearby airplanes, when the transmitter and receiver are within the communication range. The A2G links operate at 14 GHz in the Ku-band with a bandwidth of 250 MHz according to Qualcomm's  Next-Gen Air-Ground system envisaged for in-flight broadband access. For implementation simplicity, we assume that the A2A links also operates at 14 GHz as in~\cite{vondra2018integration}.

The maximum A2G and A2A link distances mainly depend on the altitude of the transmitter and receiver, because the direct visibility of two points is limited by the curvature of the Earth. For example, given a common flight altitude of 12 km, the maximum A2G link and A2A link distances are about 390 km and 780 km, respectively. The maximum A2G link distance determines the maximal coverage area on the ground/sea, i.e. the footprint, of a single airplane. For the aforementioned example, the radius of an airplane's maximum footprint can be also estimated as 390 km. Given the velocity of airplanes, such as 900 km/h (i.e., 0.25 km/s) during cruise, the maximum visibility duration of an airplane observed at ground/sea (i.e. the maximum lifetime of an A2G link) can be estimated as 780/0.25 s $\approx$ 3120 s $=$ 52 min. 
On the other hand, based on the maximum A2A link distance, it can be estimated that each airplane can be connected to dozens of other airplanes on average, given a practical flight density of $10^{-5}$ airplane/km$^2$~\cite{zhang2019aeronautical}, which makes multi-hop transmission possible for crossing the ocean.

\subsubsection{Space-Layer}
A LEO satellite constellation (SC) is considered for the space-layer due to its relatively low propagation-delay compared to GEO satellites. As an illustration, the LEO SC is deployed according to the Iridium-Next SC, which contains 66 operational LEO satellites at an altitude of 781 km.  

To support high throughput, we consider satellites using Ka-band according to the 3GPP specifications, where both the downlink (at 20 GHz) and uplink (at 30 GHz), have a bandwidth of 400 MHz~\cite{3gpp2018study}.
Each satellite can communicate with its neighboring satellites via inter-satellite links at 23~GHz, and can form 48 spot-beams down to the Earth with a frequency reuse factor of three~\cite{vondra2018integration}. As a result, the footprint radius of each spot-beam is about 200 km and of the full beam is about 2300~km. The orbital velocity of a satellite can be calculated as 7.5 km/s according to its altitude, resulting in the maximum lifetime for a satellite-to-ground/sea (S2G) link of about 10 min, which is much lower than that of the AANET.

Although the total bandwidth of LEO SC is higher than that of the A2G link, it is shared by all the air and ground/sea devices within each spot-beam. By contrast,  multiple airplanes may be available for serving the devices within a certain area. Given a practical airplane density of $10^{-5}$ airplane/km$^2$, the available bandwidth per A2G link can be twice higher than that of per S2G link.

\subsection{Routing Performance Metrics}
To support various services, multiple  metrics should be considered for characterizing the routing performance in the SAGIN. In this article, we focus on the E2E delay, E2E throughput, and path-lifetime, which are defined as follows:

\begin{enumerate}
	\item The E2E delay is defined as the summation of all the delay components along the path from the source-node (SN) to destination-node (DN). Specifically, $D_{\sf link}(i,j)$ denotes the link-delay between nodes $i$ and $j$, which is composed of the propagation-delay and transmission-delay, and  $D_{\sf que}(j)$ denotes the queuing-delay at node $j$. 
	\item The E2E throughput is defined as the achievable data-rate between the SN and DN, which is limited by the lowest-throughput link along the path, where $C(i,j)$ is the throughput between nodes $i$ and $j$, which can be calculated using Shannon's capacity formula.
	\item The path-lifetime is defined as the duration when every transmitter-receiver pair along the path has direct visibility, which is limited by the lowest link-lifetime along the path. We use $L(i,j)$ to denote the link-lifetime between nodes $i$ and $j$, which describes the duration of direct visibility between nodes $i$ and $j$.
\end{enumerate}

In the SAGIN considered, the maximum A2A link distance is shorter than the minimum S2G link. Therefore, the propagation-delay of each link in AAENT is lower than that in the LEO SC. However, since the AANET normally requires multiple hops for covering a long distance, say in the transatlantic scenario, it may result in longer accumulated queuing-delay, and hence increases the overall E2E delay. 

On the other hand, due to the shorter link distance of AANETs and owing to their potentially higher available bandwidth per user device, the E2E throughput of AANET may be higher than that of LEO SC. Moreover, again, the velocity of an airplane is much lower than that of a LEO satellite, hence its path-lifetime is potentially longer than that of LEO SC, which may improve its routing stability. Nevertheless, the airplane density fluctuates during a day due to the flight schedules. Consequently, the coverage of AANET is time-varying. By contrast, LEO SC is capable of providing 24-hour global coverage. Therefore, it is worth investigating the integration of AANET and LEO SC for combining the benefits of both networks.

\section{Deep Learning Aided Routing}
For ease of exposition, we commence from a single-objective routing problem minimizing the E2E delay, and introduce the corresponding DL-aided routing algorithm. Then, we extend it to a challenging multi-objective scenario where the E2E delay, E2E throughput, and path-lifetime are simultaneously optimized.
\subsection{Single-Objective Routing}
When the global information regarding the network topology, the queuing-delay of each node and the link-delay between every two nodes are available, the E2E delay minimization problem can be solved by classic shortest-path search algorithms by treating the summation of link-delay and queuing-delay as the ``distance", i.e. the edge weight. However, the node positions change rapidly due to the high velocity of airplanes and LEO satellites. Consequently, this may impose substantial signaling overhead for keeping the required information up-to-date in order to implement any shortest-path algorithm.  Therefore, finding the minimum-delay path in a distributed manner that solely relies on local information is desirable. 

\begin{figure}[!htb]
	\centering
	\includegraphics[width=0.45\textwidth]{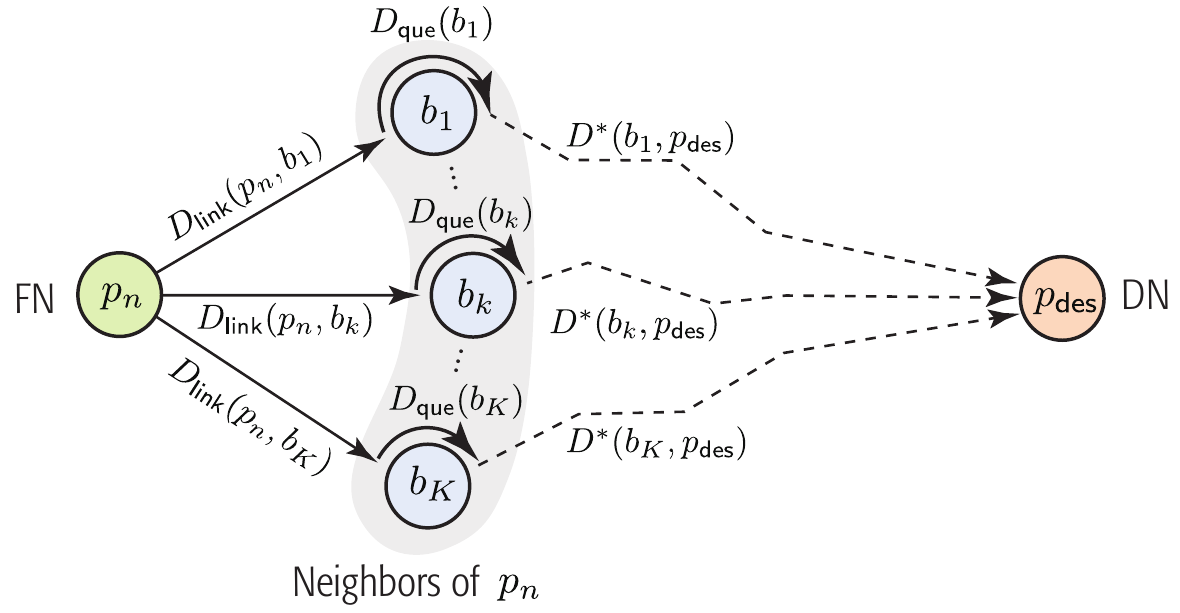}
	\caption{Optimal substructure of the minimum-delay routing problem.}
	\label{fig:opt}
\end{figure}

To commence, we revisit the minimum-delay routing problem, which has a optimal substructure shown in Fig.~\ref{fig:opt}. Assume that the packet is currently located at the forwarding node (FN) $p_n$. Then, the optimal next hop minimizing
the overall E2E delay from the current FN to the DN $p_{\sf des}$ can be characterized by
\begin{equation}
p_{n+1} = \arg\min_{b\in \mathcal{B}} \{ D_{\sf link}(p_n, b) + D_{\sf que}(b) + D^*(b, p_{\sf des})\}, \label{eqn:opt}
\end{equation}
where $\mathcal{B}$ is set of nodes that are within the communication range of the FN, i.e. the set of neighbors, and $D^*(b, p_{\sf des})$ presents the minimum delay from each neighbor $b$ to the DN.

The optimal substructure indicates that if we know the delay from the FN to each of its neighbor, and also know the minimum delay from each of its neighbors to the DN, then the optimal next hop can be
readily obtained from~\eqref{eqn:opt}. Consequently, the optimal routing path can be obtained by determining the optimal next hop one by one until the packet reaches its DN. 

\begin{figure*}[!htb]
	\centering
	\includegraphics[width=1\textwidth]{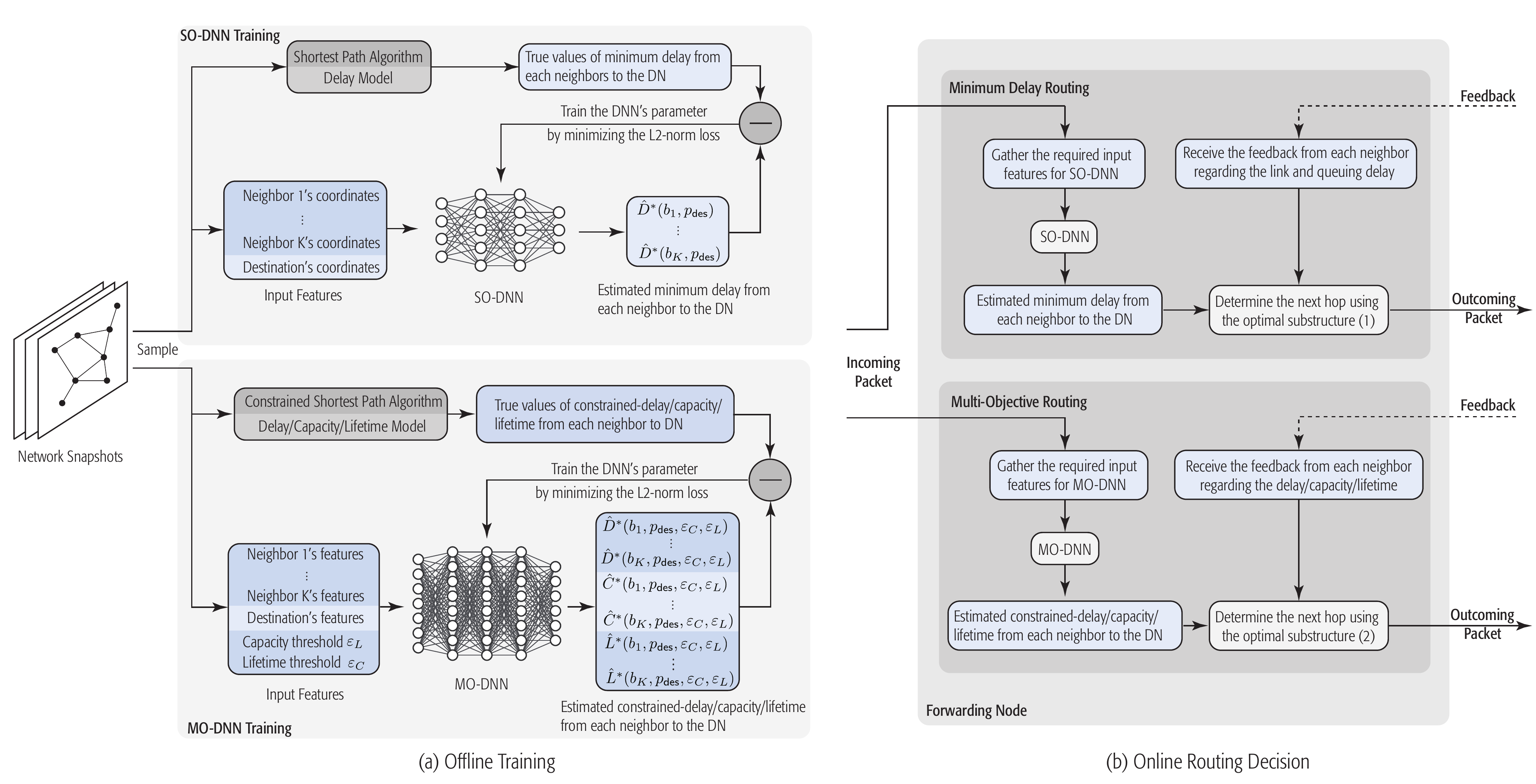}
	\caption{Offline training and online routing decision.}
	\label{fig:dnn}
\end{figure*}

In practice, the delay from the FN to each of its neighbor can be readily measured  by each neighbor and fed back to the FN almost instantly, while the minimum delay from each neighbor to the DN largely depends on  the global network topology, which is dynamic and it is not known at the FN. Fortunately, the satellite trajectories can be accurately predicted and the passenger flight trajectories are pre-planned before takeoff with fixed pattern and schedule. This suggests that the global network topology is strongly correlated with the local topology and such correlation can be learned from the quasi-predictable trajectories, which further motivates us to use a single-objective deep neural network (SO-DNN), for learning the mapping from the local geographical features to the minimum delay commencing from each of the FN's neighbors to the DN.

To reflect the local topology, the input features of the SO-DNN are designed as the coordinates (including the latitude, longitude and altitude) of the FN's neighbors and of the DN. The desired output of the SO-DNN is the minimum delay from the neighbors to the DN. 

The training procedure of the SO-DNN is shown in Fig.~\ref{fig:dnn}, where the SO-DNN is trained in a supervised manner. The goal is to minimize the loss function composed of the mean-squared-error between the actual output  and the desired output of the SO-DNN. To generate the training samples, we can retrieve the position of each airplane and each satellite at each timestamp from the pre-planned flight trajectories as well as the predicted satellite trajectories, and then create a series of snapshots regarding the whole network topology. 

For each snapshot of the network topology, the link-delay can be calculated based on the coordinate of each node and the delay model. As for the queuing-delay, since we aim for training the SO-DNN for embedding the network topology information, which is independent of the packet traffic, the queuing-delay is set to the same and constant for all the nodes during training. In this way, the total queuing-delay is actually determined by the number of hops along the path. Then, a standard shortest-path algorithm, such as the Floyd-Warshall algorithm,  can be used for computing the minimum delay between each possible source-destination pair as the training label (i.e. the desired output of the SO-DNN). In Fig.~\ref{fig:dnn}(a), we summarize the offline training phase. 

Once the SO-DNN is sufficiently well trained, it can be copied to each airplane before takeoff for supporting their online routing decisions. During the online routing decision phase, the FN obtains the required coordinates from the on-board automatic dependent surveillance-broadcast (ADS-B) system, and gathers the input features for the SO-DNN. Then, from the output of the SO-DNN, the FN can obtain the estimate of the minimum delay $\hat D^*(b, p_{\sf des})$ from each of its neighbors to the DN, as shown in Fig.~\ref{fig:dnn}(b). Finally, by substituting the estimated minimum delay together with the delays measured by and fed back from each neighbor into the optimal substructure \eqref{eqn:opt}, the next-hop can be determined. 

To improve the online adaptability further, we can recursively exploit the optimal substructure. Specifically, the minimum delay from each neighbor to the DN in~\eqref{eqn:opt} can be estimated by recursively using~\eqref{eqn:opt} instead of by directly using the SO-DNN. Then, by allowing each neighbor to estimate the minimum delay from its neighbor (i.e. the next-but-one neighbors of the FN) to the DN using the SO-DNN, and by letting each next-but-one neighbor feed back the link- and queuing-delays, additional real-time delay information can be exploited for improving the routing decisions. 

\subsection{Multi-Objective Routing}
Next, we conceive multi-objective routing for simultaneously minimizing the E2E delay, maximizing the E2E throughput and maximizing the path-lifetime. 

Typically, bio-inspired metaheuristics, such as multi-objective evolutionary algorithms (MOEAs), are harnessed for solving multi-objective optimization problems. However, MOEAs require global knowledge regarding the status (e.g., delay, throughput, lifetime) of every single possible link in the network for running the optimization, which is not feasible in large-scale SAGINs having high-dynamic network topology. Moreover, the Pareto optimality of MOEAs are not mathematically guaranteed.

Again, we resort to DL in order to exploit local information for solving the multi-objective routing problem without requiring global information. Similar to the minimum-delay routing problem, we first discover its optimal substructure. The multi-objective routing problem can be transformed into the following $\varepsilon$-constraint problem:
\begin{align}
\text{Objective:}~ &\text{Optimize routing to minimize the E2E delay} \nonumber \\
\text{Subject to:} ~&\text{E2E troughput larger than $\varepsilon_C$} \nonumber\\
& \text{and path-lifetime larger than $\varepsilon_L$}. \nonumber
\end{align}
It can be proved that all Pareto-optimal solutions can be obtained by solving the aforementioned $\varepsilon$-constraint problem with moderate complexity~\cite{chankong1983multiobjective,liu2021iot}. Furthermore, by varying the values of $\varepsilon_C$ and $\varepsilon_L$, all the Pareto-optimal solutions and hence the entire Pareto front can be found. 

Then, our goal is to solve a series of constrained minimum-delay (single-objective) routing problems. Given the network topology, the queuing-delay of each node, and the delay/throughput/lifetime of each link, the constrained minimum delay routing problem can be solved effortlessly by modifying the standard shortest-path algorithm using the following procedure:
\begin{enumerate}
	\item Convert the constrained minimum-delay routing problem into a standard shortest-path equivalent problem by deleting all the links that have a throughput no higher than $\varepsilon_C$ or have a lifetime no longer than $\varepsilon_L$.
	\item Solve the converted  shortest-path problem.
\end{enumerate}

The above procedure can be regarded as adding an infinite delay penalty to the link that violates the throughput or lifetime constraint, and then solve the converted minimum-delay problem using the standard shortest-path algorithm. Therefore, similar to \eqref{eqn:opt}, we can derive the optimal substructure for determining the next hop as follows 
\begin{align}
p_{n+1}  = &\arg\min_{b\in \mathcal{B}} \{D_{\sf link} (p_{n}, b) + D_{\sf que}(b) + D^*(b, p_{\sf des}, \varepsilon_C, \varepsilon_L)  \nonumber\\
& + \underbrace{ \lambda [\varepsilon_C - C(p_{n}, b)]^+ + \lambda[\varepsilon_C - C^*(b, p_{\sf des}, \varepsilon_C, \varepsilon_L)]^+}_{\text{Penalty for violation of throughput constraint}}  \nonumber\\
& + \underbrace{\lambda [\varepsilon_L - L(p_{n}, b)]^+ + \lambda[\varepsilon_L - L^*(b, p_{\sf des}, \varepsilon_C, \varepsilon_L)]^+}_{\text{Penalty for violation of lifetime constraint}}  \}. \label{eqn:opt2}
\end{align}
where $D^*(b, p_{\sf des}, \varepsilon_C, \varepsilon_L)$, $C^*(b, p_{\sf des}, \varepsilon_C, \varepsilon_L)$ and $L^*(b, p_{\sf des},$ $\varepsilon_C, \varepsilon_L)$ are termed as the constrained-delay, constrained-throughput, and constrained-lifetime, respectively. When there exists at least one path spanning from neighbor $b$ to the DN $p_{\sf des}$ satisfying the minimum-throughput constraint $\varepsilon_C$ and minimum-lifetime constraint $\varepsilon_L$, then the constrained-delay/throughput/lifetime represent the actual delay/throughput/lifetime achieved by the minimal delay path among the paths that satisfy the throughput and lifetime constraints. By contrast, when there is no path from $b$ to $p_{\sf des}$ that satisfies the throughput and lifetime constraint, then we opt for the delay/rate/lifetime achieved by the path from $b$ to $p_{\sf des}$ whose throughput and lifetime are closest to $\varepsilon_C$ and $\varepsilon_L$. Furthermore, $\lambda$ in \eqref{eqn:opt2} is the penalty coefficient for punishing the violation of the throughput or lifetime constraint, where $[x]^+ = \max\{x, 0\}$. The penalty term increases with the violation of E2E throughput or path-lifetime constraints, and hence the neighbors that are away from satisfying those constraints are less likely to be selected as the next hop.

In practice,   the link-delay and queuing-delay as well as the link-throughput and link-lifetime can be measured by the neighbor and fed back to the FN. Then, the optimal next-hop can be determined locally based on the optimal substructure~\eqref{eqn:opt2}, once the values of constrained-delay/throughput/lifetime of each neighbor are available at the FN.  

Therefore, similar to SO-DNN, we use a multi-objective deep neural network (MO-DNN) for learning the mapping from the local geographical features to the constrained-delay/throughput/lifetime of each neighbor. In practice, the penalty coefficient $\lambda$ is set to a limited value, because when it is excessive (e.g., when approaching infinity) the optimal substructure~\eqref{eqn:opt2} becomes extremely sensitive to the estimation error of the constrained-delay/throughput/lifetime.

The input features of the MO-DNN are designed in Fig.~\ref{fig:dnn}. Compared to the SO-DNN, the speed and heading of the current node, of its neighbors and of the DN are also included into the features together with the node's coordinates, because the constrained-lifetime depends on those parameters. Moreover, the throughput and lifetime thresholds are also included into the input feature because they affect the values of the constrained-delay/throughput/lifetime.

Similar to the training of SO-DNN, the training procedure of the MO-DNN is shown in Fig.~\ref{fig:dnn}. In contrast to SO-DNN, we also sample different values of throughput and lifetime thresholds when generating the training sample.
The online routing decision phase is illustrated in Fig~\ref{fig:dnn}(b). In contrast to single-objective routing, by varying the value of $(\varepsilon_C, \varepsilon_L)$, multiple paths leading to the DN can be discovered as the solutions of the multi-objective routing problem. Similar to DL-aided minimum-delay routing, we can also improve the online adaptability of DL-aided multi-objective routing by recursively using the optimal substructure~\eqref{eqn:opt2}, which is not detailed here for conciseness.

\section{Performance Evaluation}
In this section, we quantify the benefits of integrating AANET with LEO SC by simulations based on real satellite, flight and shipping data. 
\begin{table*}[!htb]
	\centering
	\scriptsize
	\caption{Simulation Parameters~\cite{3gpp2018study,vondra2018integration}}
	\setlength{\extrarowheight}{1.2pt}
	\begin{tabular}{|r|c|c|c|c|}
		\hline
		\textbf{Node} & \multicolumn{2}{c|}{\textbf{Satellite}} & \multicolumn{2}{c|}{\textbf{Airplane/Ship/GS}} \\
		\hline
		\textbf{Transmission mode} & To airplane/ship/GS & To satellite & To airplane/ship/GS/BS & To satellite \\ \hline
		\multirow{1.5}[2]{*}{\textbf{Carrier frequency}} & 20 GHz (Downlink)  & \multirow{1.5}[2]{*}{23 GHz} & \multirow{1.5}[2]{*}{14 GHz} & 20 GHz (Downlink)  \\
		& 30 GHz (Uplink) &       &       & 30 GHz (Uplink) \\ \hline
		\textbf{Total bandwidth} & 400 MHz & 400 MHz & 250 MHz & 400 MHz \\ \hline
		\textbf{Bandwidth per link} & 5 MHz & NA & 10 MHz & 5 MHz \\ \hline
		\textbf{Transmit power} & \multicolumn{2}{c|}{21.5 dBm} & 30 dBm & 33 dBm \\ \hline
		\textbf{Transmit antenna gain} & \multicolumn{2}{c|}{38.5 dBi} & 25 dBi & 43.2 dBi  \\ \hline
		\textbf{Receive antenna gain} & \multicolumn{2}{c|}{38.5 dBi} & 25 dBi & 39.7 dBi \\ \hline
		\multirow{1.5}[2]{*}{\textbf{Antenna gain-to-noise-temperature}} & \multicolumn{2}{c|}{\multirow{1.5}[2]{*}{13 dB/K}} & Airplane: 1.5 dB/K & Airplane: 16.2 dB/K \\
		& \multicolumn{2}{c|}{} & GS/Ship: 1.2 dB/K & GS/Ship: 15.9 dB/K \\
		\hline
	\end{tabular}%
	\label{tab:parameter}%
\end{table*}%

\subsection{Dataset}
The satellite traces  are generated based on their orbit information (more specifically, the two-line element set) of the Iridium-NEXT SC provided by the North American Aerospace Defense Command. The flight data was collected over the North Atlantic ocean on 29th of June in 2018. The status of each flight was recorded in the format of [timestamp, latitude, longitude, altitude, speed, heading] for every 10 s over the complete 24 hours of the selected date. The shipping data was also collected over the North Atlantic ocean in the same format from marinetraffic.com.

Since we only have the true flight traces during a single day, we have to generate multiple datasets of flight traces for testing outside of the training set. To reflect the mismatch between the flight traces used in training and testing, we randomly shift each true flight trace along the timeline in order to generate multiple synthetic flight datasets, respectively, used for training and testing. Specifically, the random shift is drawn from a Gaussian distribution with zero mean
and a standard deviation of 30 min. We further divide the training dataset into four time windows, each having six hours, where a MO-DNN was trained separately. In particular, for the simulations presented in this article, the MO-DNN was trained using the pre-planned flight traces and predicted satellite traces within a time
window of 12:00 -- 18:00 UTC. Once the MO-DNN is well-trained, the MO-DNN can be used for assisting the routing decisions during 12:00 -- 18:00 UTC in the testing dataset. We note that the training does not rely on ship distributions, because the MO-DNN is used for learning the constrained-delay/throughput/lifetime of each relaying neighbors that exclude the ships (i.e., packet source) as shown in Fig.~\ref{fig:dnn}(a).
\subsection{Simulation Settings}

The system configurations are listed in Table~\ref{tab:parameter}. 
To generate the training labels, the queuing-delay is set to a constant 10~ms for each node during training. By contrast, to reflect the heterogeneous traffic load of each node during testing, the queuing-delay of each node is drawn from a [0, $\infty$)-truncated Gaussian distribution with a mean of 10 ms and a standard deviation of 5 ms. The packet size is 1 KBytes and the DN is set as the GS located at the Port of Southampton in the U.K. as labeled in Fig.~\ref{fig:path}.

\subsection{Hyper-Parameters for DL-aided Routing Algorithm}
Parameter sweeps are used for tuning our DL-aided routing algorithm. The main hyper-parameters are set as follows. The MO-DNN has three hidden layers, where the first two hidden layers have 300 neurons and the third hidden layer is split into three streams each having 100 neurons, for learning the constrained-delay/throughput/lifetime, respectively. We use ReLU as the activation function for all the hidden layers, and employ batch normalization, since each
dimension of the input has different units. Adam optimizer with an initial learning rate of 0.001 is used for training the MO-DNN and the mini-batch size is 1000. The throughput threshold $\varepsilon_C$ and lifetime threshold $\varepsilon_L$ are swept over [0, 70] Mbps with a step size of 5 Mbps and [0, 30] min with a step size of 5 min, respectively, in order to discover multiple paths. The penalty coefficient $\lambda$ is set to 10. 

\subsection{Simulation Results}

In the following, we first use our modified shortest-path algorithm assuming perfect global information to find the Pareto-optimal paths for evaluating the potential of integrating AANET and LEO SC, and finally employ the proposed DL-aided multi-objective routing algorithm for finding the paths by considering a practical scenario when only local information is available. 

\begin{figure}[!htb]
	\centering
	\includegraphics[width=0.5\textwidth]{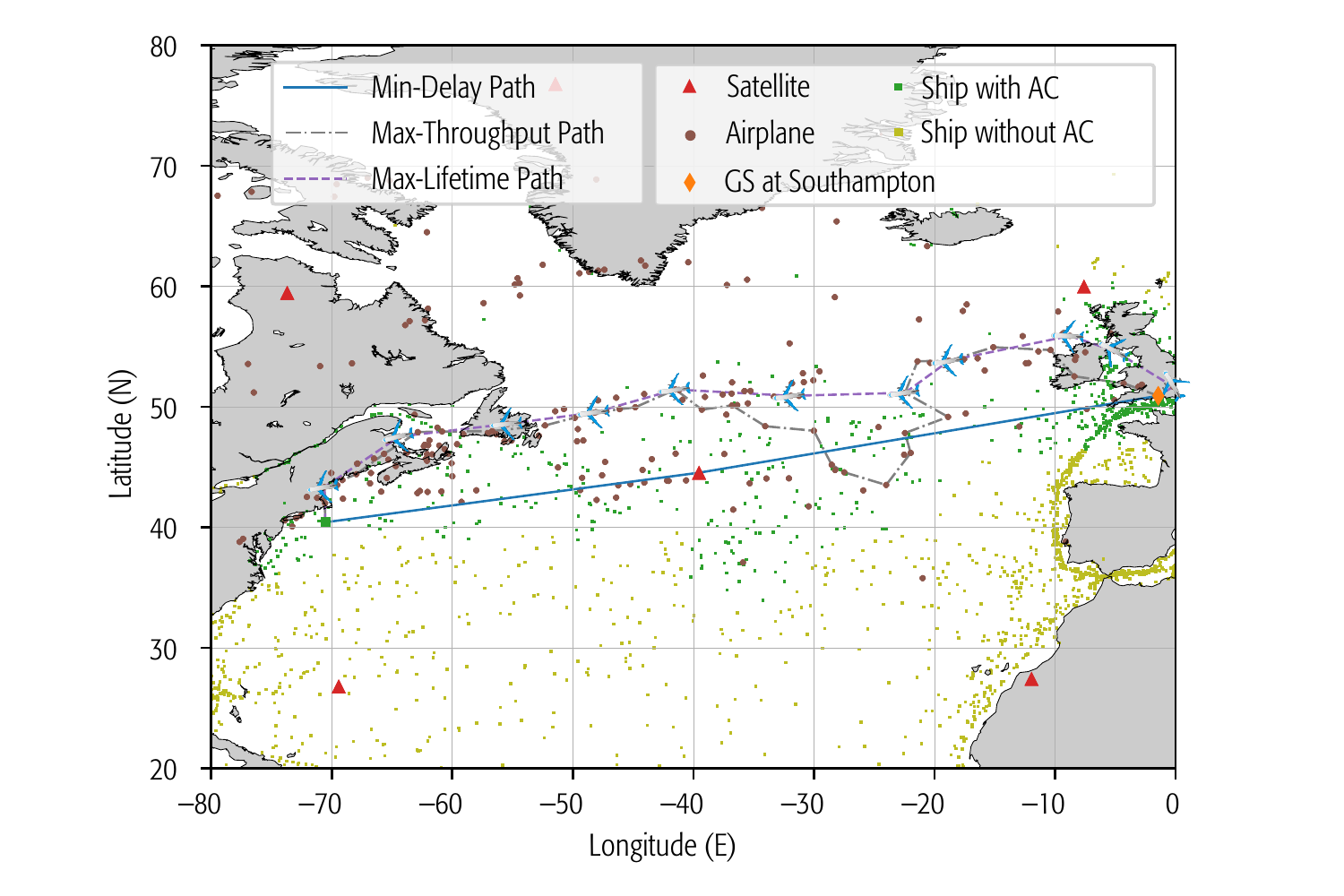}
	\caption{A snapshot of the network at 15:00 UTC, Jun. 29, 2018.}
	\label{fig:path}
\end{figure}

In Fig.~\ref{fig:path}, we illustrate a snapshot of the network at 15:00 UTC. The position of each satellite, airplane, ship and GS are marked on the projected 2D map. In particular, we use different colors to distinguish the ships that can be successfully connected to the DN solely relying on the airplanes from those that cannot (with legend ``Ship with AC" and ``Ship without AC", respectively). It can be observed that not all the ships can be covered by the AANET. 

\begin{figure*}[!htb]
	\centering
	\includegraphics[width=1\textwidth]{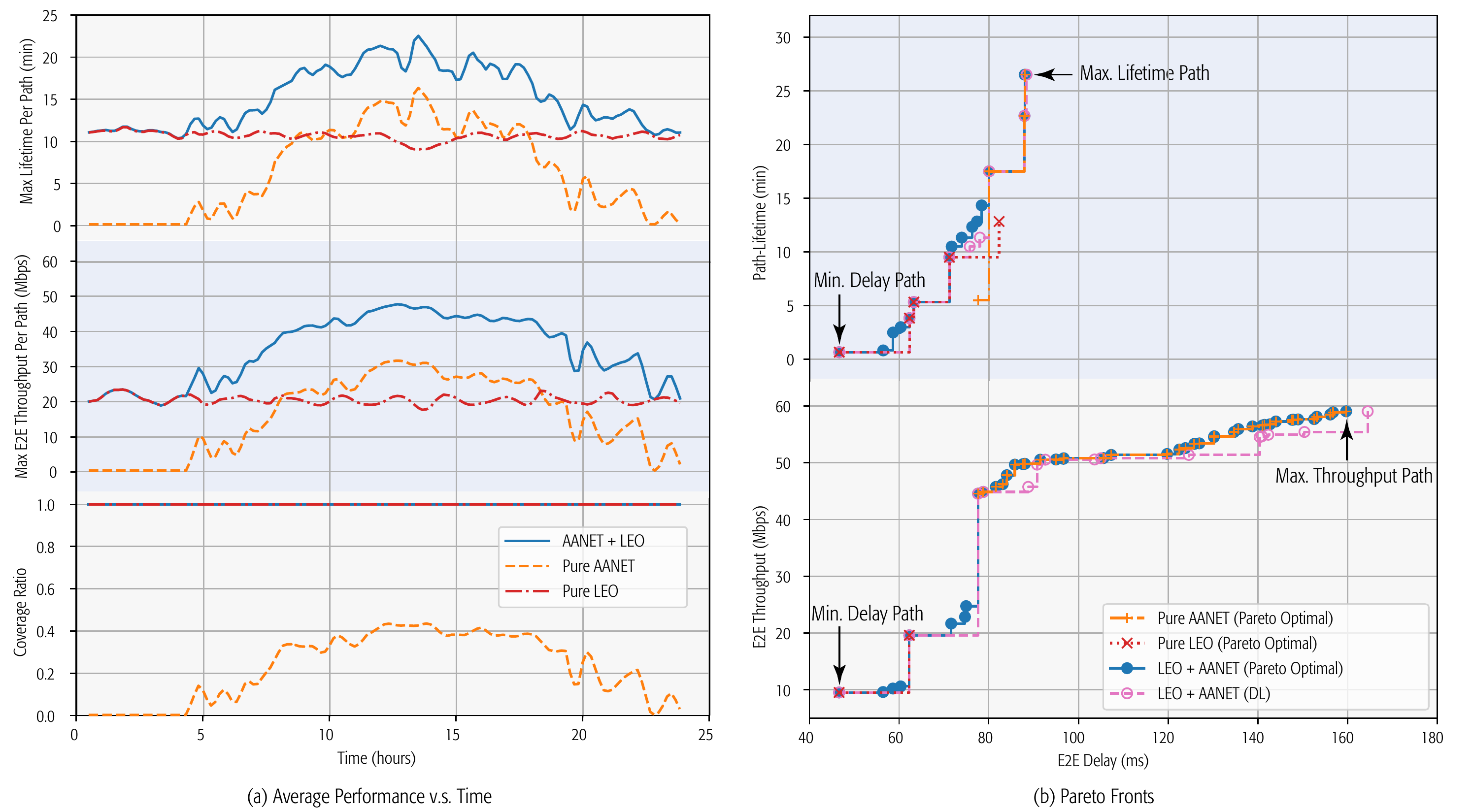}
	\caption{Simulation results, 00:00 -- 24:00 UTC, Jun. 29, 2018.}
	\label{fig:result}
\end{figure*}

Since there may exist more than one Pareto-optimal path from the SN to the DN, in Fig.~\ref{fig:path}, we further plot the minimum-delay path, maximum-throughput path, and maximum-lifetime path among all the paths found by our modified shortest-path algorithm. The SN is chosen from the north-east coast of US for analyzing the transatlantic communication scenario. We can see that the minimum-delay path is a two-hop path using a satellite as the relaying node, while the maximum-throughput path and the maximum-lifetime path are formed by airplanes. Specifically, the maximum-throughput path has the highest number of hops in order to reduce the distance of each link for maximizing the throughput of the path. We further plot the heading of the flights on the maximum-lifetime path to understand how the path is formed. We can see that the angle between the flight headings of adjacent nodes is small so that the lifetime of each link is maximized.

To characterize the merits of integrating AANET and LEO SC, we compare three systems, namely pure AANET, pure LEO SC, and integrated AANET with LEO SC in the following. We first plot the average performance of the path from each ship to the GS of the three systems during a whole day as shown in Fig.~\ref{fig:result}(a). It can be observed that the percentage of ships that can be connected to the GS solely relying on the AANET fluctuates during the day, because the coverage of AANET depends on the flight density, which varies during a whole day. The maximum coverage ratio of AANET can be as high as 40\% lasting about 8 hours. By contrast, LEO SC alone or combining LEO SC and AANET together can provide a 100\% coverage for the entire 24 hours. As for the E2E throughput and path-lifetime, we can see that pure AANET achieves an average maximum E2E throughput and path-lifetime that are similar to or even higher than that of LEO SC, although the coverage ratio of AANET is less than half of that of LEO SC. When integrating AANET with LEO SC, both the maximum E2E throughput and maximum path-lifetime can be improved up to 200\% on average\footnote{When there is no path between the SN and DN, the E2E throughput and path-lifetime are set to zeros.}.

To analyze the optimal tradeoff between multiple performance metrics, we respectively consider a pair of twin-objective routing problems for a clearer visualization. Specifically, we first minimize the E2E delay and maximize the E2E throughput simultaneously, and then minimize the E2E delay and maximize the path-lifetime simultaneously. In Fig.~\ref{fig:result}(b), we plot the corresponding Pareto-front of the paths between the SN and DN illustrated in Fig.~\ref{fig:path}. Observe that LEO SC is more suitable for reducing the E2E delay in the transatlantic scenario due to its lower numbers of hops, while the AANET is more suitable for increasing the E2E throughput and the path-lifetime, as a benefit of its lower link distance and longer duration of visibility. By integrating AANET and LEO SC, the Pareto front can be improved and the route can be selected more flexibly among the three metrics. When only relying on the local information, our proposed DL-aided multi-objective routing algorithm, with the legend ``AANET + LEO (DL)", can discover paths that achieve near-Pareto-optimal performance. By combining the E2E delay, E2E throughput and path-lifetime, we can further obtain and visualize a 3D-pareto front~\cite{liu2021iot}.

\section{Concluding Remarks}
In this article, we proposed a SAGIN framework that integrates AANETs formed by commercial passenger airplanes and LEO SC for connecting the ships in the North Atlantic Ocean. To satisfy the requirements of heterogeneous services and adapt to the dynamics of SAGIN, we further proposed a DL-aided multi-objective routing algorithm, which exploits the quasi-predictable network topology and operates in a distributed manner. Our simulation results based on real satellite, flight and shipping data showed that the integrated network  achieves better coverage, lower E2E delay, higher E2E throughput as well as higher path-lifetime, and demonstrated that our DL-aided multi-objective routing algorithm is capable of achieving near-Pareto-optimal performance.
\bibliographystyle{IEEEtran}
\bibliography{ref}

\begin{thebibliography}{10}
\providecommand{\url}[1]{#1}
\csname url@samestyle\endcsname
\providecommand{\newblock}{\relax}
\providecommand{\bibinfo}[2]{#2}
\providecommand{\BIBentrySTDinterwordspacing}{\spaceskip=0pt\relax}
\providecommand{\BIBentryALTinterwordstretchfactor}{4}
\providecommand{\BIBentryALTinterwordspacing}{\spaceskip=\fontdimen2\font plus
\BIBentryALTinterwordstretchfactor\fontdimen3\font minus
  \fontdimen4\font\relax}
\providecommand{\BIBforeignlanguage}[2]{{%
\expandafter\ifx\csname l@#1\endcsname\relax
\typeout{** WARNING: IEEEtran.bst: No hyphenation pattern has been}%
\typeout{** loaded for the language `#1'. Using the pattern for}%
\typeout{** the default language instead.}%
\else
\language=\csname l@#1\endcsname
\fi
#2}}
\providecommand{\BIBdecl}{\relax}
\BIBdecl

\bibitem{wei2021hybrid}
T.~Wei, W.~Feng, Y.~Chen, C.-X. Wang, N.~Ge, and J.~Lu, ``Hybrid
  satellite-terrestrial communication networks for the maritime {Internet of
  Things}: key technologies, opportunities, and challenges,'' \emph{IEEE
  Internet Things J.}, vol.~8, no.~11, pp. 8910--8934, Jun. 2021.

\bibitem{vondra2018integration}
M.~Vondra, M.~Ozger, D.~Schupke, and C.~Cavdar, ``Integration of satellite and
  aerial communications for heterogeneous flying vehicles,'' \emph{IEEE Netw.},
  vol.~32, no.~5, pp. 62--69, Sept./Oct. 2018.

\bibitem{zhang2019aeronautical}
J.~Zhang, T.~Chen, S.~Zhong, J.~Wang, W.~Zhang, X.~Zuo, R.~G. Maunder, and
  L.~Hanzo, ``Aeronautical ad hoc networking for the
  {I}nternet-above-the-clouds,'' \emph{Proc. IEEE}, vol. 107, no.~5, pp.
  868--911, May 2019.

\bibitem{huang2019airplane}
X.~Huang, J.~A. Zhang, R.~P. Liu, Y.~J. Guo, and L.~Hanzo, ``Airplane-aided
  integrated networking for {6G} wireless: Will it work?'' \emph{IEEE Veh.
  Technol. Mag.}, vol.~14, no.~3, pp. 84--91, Sept. 2019.

\bibitem{liu2018space}
J.~Liu, Y.~Shi, Z.~M. Fadlullah, and N.~Kato, ``Space-air-ground integrated
  network: A survey,'' \emph{IEEE Commun. Surveys Tuts.}, vol.~20, no.~4, pp.
  2714--2741, 4th Quart. 2018.

\bibitem{luong2019applications}
N.~C. Luong, D.~T. Hoang, S.~Gong, D.~Niyato, P.~Wang, Y.-C. Liang, and D.~I.
  Kim, ``Applications of deep reinforcement learning in communications and
  networking: A survey,'' \emph{IEEE Commun. Surveys Tuts.}, vol.~21, no.~4,
  pp. 3133--3174, 4th Quart. 2019.

\bibitem{liu2020optimizing}
D.~Liu, C.~Sun, C.~Yang, and L.~Hanzo, ``Optimizing wireless systems using
  unsupervised and reinforced-unsupervised deep learning,'' \emph{IEEE Netw.},
  vol.~34, no.~4, pp. 270--277, Jul./Aug. 2020.

\bibitem{yang2020ai}
T.~Yang, J.~Chen, and N.~Zhang, ``{AI}-empowered maritime {Internet of Things}:
  A parallel-network-driven approach,'' \emph{IEEE Netw.}, vol.~34, no.~5, pp.
  54--59, Sept./Oct. 2020.

\bibitem{kato2019optimizing}
N.~Kato, Z.~M. Fadlullah, F.~Tang, B.~Mao, S.~Tani, A.~Okamura, and J.~Liu,
  ``Optimizing space-air-ground integrated networks by artificial
  intelligence,'' \emph{IEEE Wireless Commun.}, vol.~26, no.~4, pp. 140--147,
  Aug. 2019.

\bibitem{liu2021deep}
D.~Liu, J.~Cui, J.~Zhang, C.~Yang, and L.~Hanzo, ``Deep reinforcement learning
  aided packet-routing for aeronautical ad-hoc networks formed by passenger
  planes,'' \emph{IEEE Trans. on Veh. Techn.}, vol.~70, no.~5, pp. 5166--5171,
  May 2021.

\bibitem{jingjing}
J.~Cui, S.~X. Ng, D.~Liu, J.~Zhang, A.~Nallanathan, and L.~Hanzo,
  ``Multiobjective optimization for integrated ground-air-space networks,''
  \emph{IEEE Veh. Technol. Mag.}, vol.~16, no.~3, Sept. 2021.

\bibitem{li2020enabling}
X.~Li, W.~Feng, J.~Wang, Y.~Chen, N.~Ge, and C.-X. Wang, ``Enabling {5G} on the
  ocean: A hybrid satellite-{UAV}-terrestrial network solution,'' \emph{IEEE
  Wireless Commun.}, vol.~27, no.~6, pp. 116--121, Dec. 2020.

\bibitem{3gpp2018study}
3GPP, ``Solutions for {NR} to support non-terrestrial networks ({NTN})
  ({R}elease 16),'' TR 38.811, Dec. 2019.

\bibitem{chankong1983multiobjective}
V.~Chankong and Y.~Y. Haimes, \emph{Multiobjective decision making: theory and
  methodology}.\hskip 1em plus 0.5em minus 0.4em\relax New York: Elsevier
  Science, 1983.

\bibitem{liu2021iot}
D.~Liu, J.~Zhang, J.~Cui, S.-X. Ng, R.~G. Maunder, and L.~Hanzo, ``Deep
  learning aided packet routing in aeronautical ad-hoc networks relying on real
  flight data: From single-objective to near-pareto multi-objective
  optimization,'' \emph{IEEE Internet Things J.}, Early Access, 2021.

\end{thebibliography}

\end{document}